\documentclass{article}
        
        \usepackage[preprint]{paper_2026}

        \usepackage[utf8]{inputenc} 
        \usepackage[T1]{fontenc}    
        \usepackage{hyperref}       
        \usepackage{url}            
        \usepackage{booktabs}       
        \usepackage{amsfonts}       
        \usepackage{nicefrac}       
        \usepackage{microtype}      
        \usepackage{xcolor}         
        \usepackage{graphicx}
        \usepackage{algorithm}
        \usepackage{amsmath}
        \usepackage{amssymb}
        \usepackage{mathtools}
        \usepackage{amsthm}
        \usepackage{graphicx}
        \usepackage{subcaption}

        \usepackage{algorithm}
        \usepackage{amsmath}
        \usepackage{amssymb}
        \usepackage{mathtools}
        \usepackage{amsthm}
        \usepackage{graphicx}
        \usepackage{subcaption}
        \usepackage{hyperref}
        \usepackage{url}
        \usepackage{graphicx}
        \usepackage{enumitem}
        \usepackage{hyperref}       
        \usepackage{url}            
        \usepackage{booktabs}       
        \usepackage{amsfonts}       
        \usepackage{nicefrac}       
        \usepackage{microtype}      
        \usepackage[table]{xcolor}
        \usepackage{amsmath}    
        \usepackage{amssymb}   
        \usepackage{bm}   
        \usepackage[most]{tcolorbox}
        \usepackage{subcaption}
        \usepackage{amsthm}
        \usepackage{amssymb}
        \usepackage{wrapfig}
        \usepackage{xcolor}
        \usepackage{tabularx}
        \usepackage{array}       
        \usepackage{makecell}    
        \usepackage{color}
        \usepackage{siunitx}
        
        \usepackage[ruled,vlined,linesnumbered,algo2e]{algorithm2e}
        
        \usepackage[capitalize,noabbrev]{cleveref}
        
        \theoremstyle{plain}

        \theoremstyle{definition}

        \theoremstyle{remark}

        \title{Efficient Scaling of LLM Training with Flexible Context Parallelism}

        \author{%
          \textbf{Yifan Niu$^{1}$}\thanks{Equal contribution.} \quad
          \textbf{Han Xiao$^{1}$}\footnotemark[1] \quad
          \textbf{Dongyi Liu$^{1}$} \quad
          \textbf{Wei Zhou$^{2}$} \quad
          \textbf{Jia Li$^{1,3}$}\thanks{Corresponding author: Jia Li (\texttt{jialee@ust.hk}).}
          \\
          $^{1}$The Hong Kong University of Science and Technology (Guangzhou)
          \\
          $^{2}$Huawei Technologies Co., Ltd.
          \\
          $^{3}$The Hong Kong University of Science and Technology
        }
        
\begin{document}

        \maketitle

        \begin{abstract}
          Scaling long-context capabilities is crucial for Large Language Models (LLMs). However, real-world data contain a large number of sequences with heterogeneous lengths. Existing training libraries for LLMs rely on static parallelism strategies, which suffer from severe load imbalance, redundant communication, and suboptimal hardware utilization under data heterogeneity. In this work, we propose \textbf{F}lexible \textbf{C}ontext \textbf{P}arallelism (FCP), an efficient parallelism strategy that adaptively reconfigures communication groups and context parallelism degrees during LLM training. We generalize more flexible non-power-of-two parallelism degrees and develop a polynomial-time algorithm to generate near-optimal parallelism strategies with only millisecond-level overhead per training batch. FCP is able to maintain high hardware efficiency even under extreme data heterogeneity. Experimental results demonstrate that FCP significantly outperforms Megatron-LM and DeepSpeed in both LLM and MLLM training, achieving up to 1.46 $\times$ speedup in average throughput while maintaining near-linear scaling efficiency across large-scale clusters. For extremely unbalanced batches, FCP even achieves 2.24 $\times$ speedup. The code is available at \href{https://github.com/ivanniu/FCP}{https://github.com/ivanniu/FCP}.
        \end{abstract}
        
        \section{Introduction}
        
        In recent years, Large Language Models (LLMs) have demonstrated remarkable success in achieving sophisticated natural language understanding and generation across a wide range of tasks~\cite{xu2025qwen25omnitechnicalreport,he2025skyworkopenreasoner1,bai2025longbenchv2deeperunderstanding,yang2025qwen3technicalreport,yang2025kimidevagentlesstrainingskill,do2026pi2structureoriginatedreasoningdata}. The impressive performance of LLMs is primarily attributed to the expansion of model parameters, the curation of massive-scale high-quality text corpora, and the refinement of training objectives and architectural designs, all of which align with the scaling laws. For instance, industry-leading models such as OpenAI’s GPT-5~\cite{singh2025openaigpt5card} and Google’s Gemini 3 Pro~\cite{gemini-3-pro} exhibit unprecedented logic, coding proficiency, and complex instruction-following capabilities. Simultaneously, open-source models like DeepSeek-V4~\cite{deepseekai2026deepseekv4} and GLM-5~\cite{glm5team2026glm5vibecodingagentic} have pushed the boundaries of reasoning performance and efficient large-scale inference.

        A fundamental challenge in scaling LLMs to ultra-long contexts is the quadratic complexity $O(L^2)$ of self-attention regarding the number of tokens~\cite{liu2023ringattentionblockwisetransformers,munkhdalai2024leavecontextbehindefficient,shao2025holitom,shao2025tokenstalkmuchsurvey}. 
        Existing parallel training strategies are broadly divided into two categories: static parallelism and dynamic parallelism. Static parallelism (e.g., Megatron-LM~\cite{narayanan2021efficient}, DeepSpeed~\cite{Deepspeed}) typically employs predefined 4D parallelism: Data Parallelism (DP)~\cite{DeanCMCDLMRSTYN12,DBLP:pytorch-ddp}, Tensor Parallelism (TP)~\cite{shoeybi2020megatronlmtrainingmultibillionparameter}, Pipeline Parallelism (PP)~\cite{GPipe,PipeDream,Narayanan2020MemoryEfficientPD}, Sequence Parallelism (SP)~\cite{brandon2023stripedattentionfasterring,li2023sequence,DeepSpeed-Ulysses} or Context Parallelism (CP)~\cite{brandon2023stripedattentionfasterring,li2023lightseq}  to distribute the training on massive clusters. However, static methods often suffer from workload imbalances and redundant communication when dealing with variable-length sequences or heterogeneous batching. To address the issue, recent research has introduced dynamic parallelism. ByteScale~\cite{ByteScale} proposes a heuristic-based sequence scheduling strategy, while FlexSP~\cite{FlexSP} optimizes data partitioning by solving a time-consuming programming problem and restricts the communication group size to powers of two, which narrows the feasible space for sequence allocation.
        
        In this work, we follow the trajectory of dynamic parallelism and rethink the essential properties that define an optimal scheduling strategy. A robust dynamic parallelism should satisfy three core requirements: (1) \textbf{Workload Balance}. The computational load across different communication groups should be balanced to minimize synchronization stalls and maximize hardware efficiency. (2) \textbf{Elastic Parallelism Degree}. The strategy should support flexible, non-power-of-two parallelism degrees to prevent short sequences from incurring redundant communication overhead, thereby adapting to highly skewed data distributions. (3) \textbf{Minimal Scheduling Overhead}. The latency of generating an assignment strategy should be sufficiently low to be fully overlapped with the computation of sequences of any length, ensuring that the scheduling process itself does not become a new computation bottleneck. 
        
        Motivated by these principles, we propose Flexible
        Context Parallelism (FCP), an efficient and adaptive parallel strategy designed to accelerate LLM long-context training on heterogeneous data. FCP dynamically reconfigures communication groups and parallelism degrees for each micro-batch to handle the skewed distributions of sequence lengths. FCP is built on Ring-style Context Parallelism (CP), allowing for arbitrary integer parallelism degrees that better fit the varying sequence lengths. To solve the NP-hard scheduling problem without introducing training latency, we develop a polynomial-time algorithm based on 2D dynamic programming to obtain approximately optimal sequence assignment solutions. We implement FCP as a non-intrusive, asynchronous module atop Megatron-LM, ensuring that scheduling overhead is fully hidden behind the computation. Furthermore, FCP can also be adapted to the training of Multimodal Large Language Models (MLLMs)~\cite{feng2025optimusacceleratinglargescalemultimodal,Zhang_2025,bai2025qwen3vltechnicalreport}. Experimental results demonstrate that FCP significantly outperforms Megatron-LM and DeepSpeed in both LLM and MLLM training, achieving up to 1.47 $\times$ speedup in average throughput while maintaining near-linear scaling efficiency across large-scale clusters. For extremely unbalanced batches, FCP even achieves 2.24 $\times$ speedup.

        \section{Related Work} 
        \textbf{Parallelism for Large-Scale Training.}
        Distributed training on massive clusters is indispensable for large-scale model training. Existing parallelism strategies can be broadly categorized into static parallelism and dynamic parallelism. Static parallelism (e.g., Megatron-LM~\cite{narayanan2021efficient}, DeepSpeed~\cite{Deepspeed}) typically employs predefined 4D parallelism: Data Parallelism (DP)~\cite{DeanCMCDLMRSTYN12,DBLP:pytorch-ddp} replicates model weights across workers and partitions data; Tensor Parallelism (TP)~\cite{shoeybi2020megatronlmtrainingmultibillionparameter} partitions weight matrices within layers; Pipeline Parallelism (PP)~\cite{GPipe,PipeDream,Narayanan2020MemoryEfficientPD} distributes model layers across different stages and processes micro-batches in a pipeline; Sequence Parallelism (SP)~\cite{brandon2023stripedattentionfasterring,li2023sequence,DeepSpeed-Ulysses} or Context Parallelism (CP)~\cite{brandon2023stripedattentionfasterring,li2023lightseq} partitions along the sequence dimension to support long-context training. However, static methods often suffer from severe workload imbalances and exhibit poor adaptability to heterogeneous sequence lengths. To address these issues, recent research has introduced dynamic parallelism. ByteScale~\cite{ByteScale} employs data-aware sharding and dynamic communication to eliminate redundant communication for short sequences. The heuristic-based scheduler relies on a greedy approximation, which may lead to sub-optimal load balancing. FlexSP~\cite{FlexSP} adjusts partitioning by solving the time-consuming Programming problem, which takes $5$-$15$ seconds per batch. FlexSP is built on DeepSpeed-Ulysses~\cite{DeepSpeed-Ulysses}, where the parallelism degree is limited to a power of two, and narrows the feasible space for heterogeneous sequence allocation.

        \section{Preliminaries} 
        
        \subsection{Large Language Models}

        The primary computational workload of Large Language Models (LLMs) resides in a stack of Transformer blocks. Each block is composed of two main components: a Multi-Head Self-Attention (MHA) module and a Feed-Forward Network (FFN). While the FFN exhibits linear complexity with respect to the sequence length $L$, the MHA module presents a fundamental scaling bottleneck. Formally, for an input hidden state $H \in \mathbb{R}^{L \times D}$, the self-attention operation is computed as:
        \begin{equation}
        \text{Attn}(Q, K, V) = \text{softmax}\left(\frac{QK^T}{\sqrt{d_k}}\right)V
        \end{equation}
        where $Q, K, V$ are linear projections of $H$. The computation of the affinity matrix $QK^T$ requires $\mathcal{O}(L^2 D)$ floating-point operations (FLOPs). As LLMs scale to ultra-long contexts, this quadratic complexity $\mathcal{O}(L^2)$ dominates the execution time per transformer layer.

        \subsection{Distributed Training Paradigms}
        \textbf{Data Parallelism.} Data Parallelism (DP)~\cite{DBLP:pytorch-ddp} partitions data along the batch dimension. Each device maintains a full replica of model parameters and synchronizes gradients via All-Reduce operations. To mitigate the redundancy of model states, Sharded Data Parallelism (SDP) (e.g., ZeRO~\cite{ZeRO2020}, FSDP~\cite{FSDP2023}) partitions model parameters, gradients, and optimizer states across devices, trading increased communication for significantly reduced per-device memory consumption.
        
        \textbf{Model Parallelism.} Model Parallelism decomposes model parameters across devices, categorized into Tensor Parallelism (TP) and Pipeline Parallelism (PP). TP~\cite{shoeybi2020megatronlmtrainingmultibillionparameter} partitions individual operators, requiring frequent synchronization of activations. Therefore, TP is typically restricted to intra-node deployment to leverage high-bandwidth interconnects. PP~\cite{GPipe,PipeDream} partitions the model by layers and only transfers boundary activations between stages, making it suitable for cross-node scaling.
        
        \textbf{Sequence-Dimension Partitioning (CP \& SP).} Both Sequence Parallelism (SP) and Context Parallelism (CP) distribute data along the sequence dimension to overcome memory bottlenecks. Sequence Parallelism (SP), such as DeepSpeed-Ulysses~\cite{DeepSpeed-Ulysses}, uses all-to-all primitives to redistribute data so that each device computes attention for a subset of heads over the full sequence.  The degree of SP must be a divisor of the number of attention heads while simultaneously fully utilizing highly optimized collective communication operators; this usually confines the parallelism degree to powers of two and limits its flexibility. Context Parallelism (CP) (e.g., Ring-Attention~\cite{brandon2023stripedattentionfasterring,li2023sequence,li2024distflashattndistributedmemoryefficientattention}) decomposes the attention computation itself. CP often employs ring-based point-to-point communication to exchange Key/Value blocks, allowing the overlap of communication with computation. Thus, CP usually supports longer context without being constrained by head-count divisibility.

        \section{Method}
        \subsection{Motivation and Observation}
        
\begin{figure*}[h]
          \vspace{-0.2cm}
          \centering
          \begin{minipage}[c]{0.49\textwidth}
            \centering
            \includegraphics[width=0.98\linewidth]{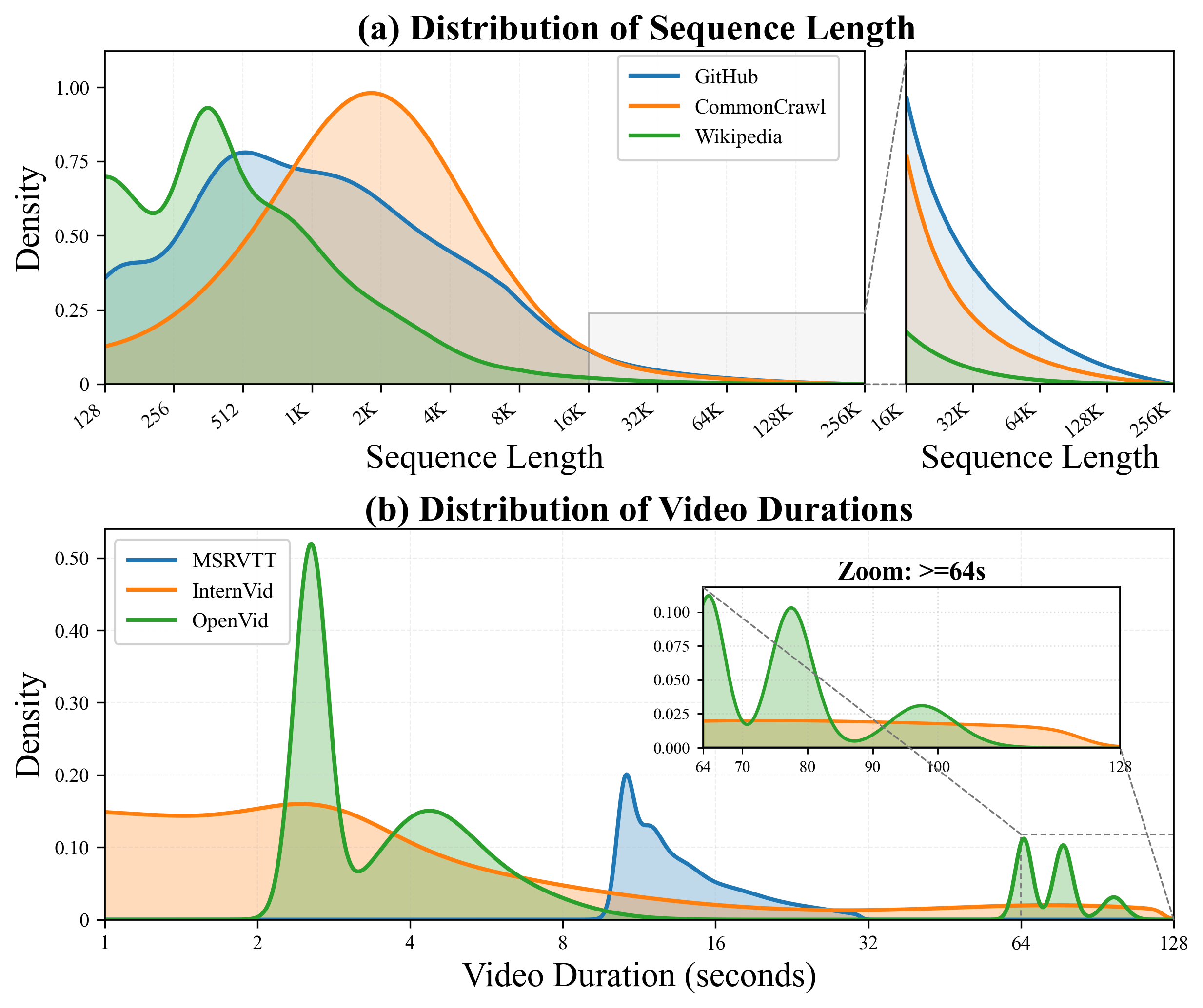}
            \vspace{-0.0cm}
            \centerline{\small (a) Data length distributions.}
          \end{minipage}
          \hfill
          \begin{minipage}[c]{0.49\textwidth}
            \centering
            \includegraphics[width=0.98\linewidth]{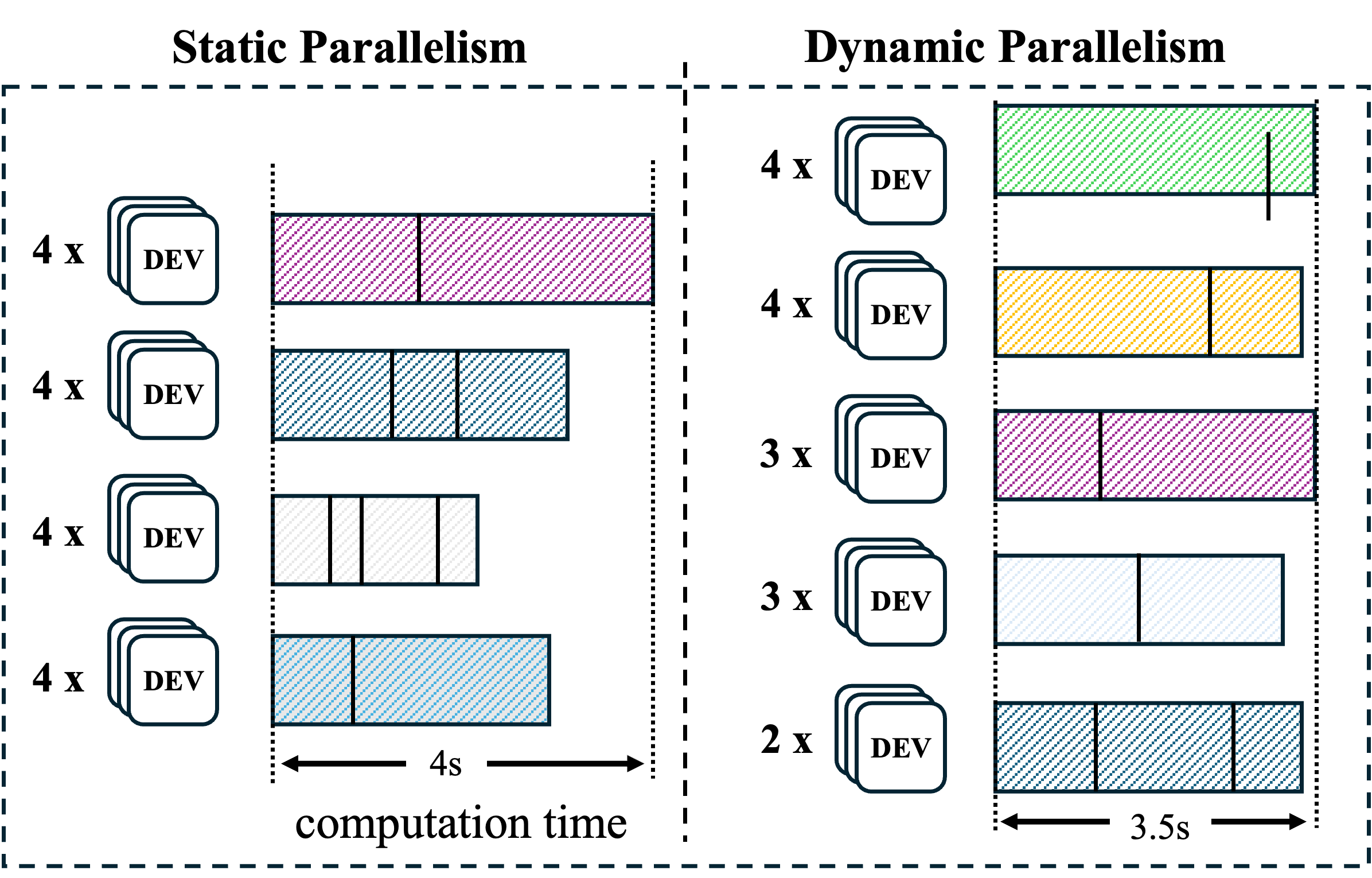}
            \vspace{-0.0cm}
            \centerline{\small (b) Static parallelism vs. dynamic parallelism.}
          \end{minipage}
          \caption{(a) Length distribution of text and multimodal datasets. (b) Compared to static parallelism, dynamic parallelism facilitates load balancing against heterogeneous data. }
          \label{fig:motivation_observation}
          \label{fig:data_dist}
          \label{fig:Static_Dynamic}
          \vspace{-0.2cm}
        \end{figure*}

        The length distributions of real-world data are skewed, exhibiting significant variations across different datasets and modalities. As shown in Figure ~\ref{fig:motivation_observation} (a), most sequences in common text datasets (e.g., GitHub~\cite{kocetkov2022stack3tbpermissively}, CommonCrawl~\cite{raffel2023exploringlimitstransferlearning}, and Wikipedia~\cite{wikidump}) and multimodal datasets (e.g., MSRVTT~\cite{xu2016msrvtt}, InternVid~\cite{wang2024internvidlargescalevideotextdataset}, and OpenVid~\cite{nan2025openvidm}) are relatively short, but a few samples are several times longer than the average. Furthermore, the length distributions of these datasets also show significant differences in mean, central location, and long-tail patterns.
        
        When training on a mix of short and long contexts, static methods often partition resources by equal token counts or fixed parallelism degree, but this does not account for the non-linear computation induced by attention. A rank processing a single long sequence $L$ performs $O(L^2)$ computation, whereas a rank processing multiple packed short sequences $\{l_i\}$ only executes $\sum O(l_i^2)$. Since $L^2 \gg \sum l_i^2$, ranks assigned long text sequences or long video-derived visual contexts become stragglers, forcing other ranks to idle at synchronization barriers, as illustrated in Figure~\ref{fig:motivation_observation} (b). Conversely, dynamic parallelism adaptively reconfigures communication groups and sequence assignment. By avoiding over-packing of short sequences and balancing the computational workload, it minimizes hardware idle time and reduces the total computation time.

        \subsection{Problem Formulation} \label{subsubsec:problem_form}

        \textbf{Notation.} In this work, we focus on the dynamic construction of CP groups, as the ring attention overlaps the communication cost well for long sequences. By dynamically assigning each sequence $s_k$ in a micro-batch $\mathcal{B}$ to a specific CP group $C_p$ with an appropriate parallelism degree $d_p$, our formulation implicitly incorporates DP. We exclude Model Parallelism (TP \& PP) from our dynamic optimization space, as the overhead of reconfiguring TP/PP groups (e.g., weight reshuffling) during runtime is unacceptable. Consequently, we treat TP and PP as predefined static configurations. In our discussion, a single ``rank" $r_n$ represents a complete model replica (i.e., TP $\times$ PP physical devices), and $N$ denotes the total number of such replicas available in the cluster.
        
        Formally, let $\mathcal{R}$ be the set of $N$ available computing units in the cluster.  We aim to dynamically partition $\mathcal{R}$ into disjoint CP groups $\mathcal{C} = \{C_p\}^P$ with degree $d_p$, and $\bigcup C_p = \mathcal{R}$. 
        Given a micro-batch of $K$ sequences $\{s_k\}$ with heterogeneous lengths and a per-rank memory budget $E$, we aim to determine: (1) the total number of CP groups, (2) the parallelism degree of each group, and (3) the assignment of each sequence to a specific CP group. 
        
        We extend the optimization objective in FlexSP to \emph{any positive integer parallelism degree} by incorporating Ring-style CP~\cite{brandon2023stripedattentionfasterring}. FlexSP restricts the parallelism degree to powers of two, limiting the feasible space for optimal configurations. Our relaxation enables finer-grained resource allocation and better adaptation to heterogeneous sequence lengths. We define the sequence assignment matrix $\boldsymbol{A} \in \{0,1\}^{K\times P}$, where $A_{k,p}=1$ represents that $s_k$ is assigned to $C_p$.  The more flexible CP group and sequence assignment problem can be formulated as follows: 
        \begin{align}
        & \quad \arg\min_{\boldsymbol{A},\mathcal{C}} \ \max \mathcal{T}(C_p), \label{eq:minimize_time} \\
        \text{s.t.} \quad & \mathcal{M}(C_p) \leq E \cdot d_{p}, \ \forall p \in [1,P] \label{cond:gen_memory} \\
        & \sum\nolimits_k{A_{k,p}} \leq K, \ \forall p \in [1,P] \label{cond:gen_mp} \\
        & \sum\nolimits_p{A_{k,p}} = 1, \ \forall k \in [1,K] \label{cond:gen_assign} \\
        & \sum\nolimits_p{d_p} \leq N \label{cond:gen_device}
        \end{align}
        where $\mathcal{T}(\cdot)$ and $\mathcal{M}(\cdot)$ denotes the execution time and the memory cost in CP group $C_p$. The optimization objective is to minimize the makespan, defined as the maximum execution time across all CP groups. The constraints are defined as follows: Cond.~\eqref{cond:gen_memory} enforces the hardware memory limit on each rank; Cond.~\eqref{cond:gen_mp} and \eqref{cond:gen_assign} guarantee the exclusive assignment of each sequence to exactly one active CP group; finally, Cond.~\eqref{cond:gen_device} ensures that the aggregate parallelism degree across all selected groups does not exceed the total rank budget $N$. Note that this optimization problem is \emph{NP-hard}.

        \subsection{Cost Estimation} 
         \textbf{Memory Estimation.} The memory consumption consists of two primary components: model states and activations. For model states, the model state memory $\text{M}_{ms}$ remains constant per rank. For activations, CP partitions each sequence $s_k$ evenly into $d_p$ chunks along the sequence dimension.  Thus, for a CP group $C_p$, the memory consumption is estimated as:
        \begin{equation}
         \mathcal{M}(C_p) = \sum_{k} A_{k,p} |s_k| \cdot \text{M}_{token} + \text{M}_{ms}, \label{eq:ring_memory_cost}\end{equation}
        where $\text{M}_{token}$ represents the activation memory per token.

        \textbf{Time Estimation.} (1) \textbf{Computation Cost.} 
        Computation cost should take MHA and FFN into consideration. To achieve a unified estimation for both LLM and MLLM workloads, we generalize the computation cost model to accommodate diverse attention mask. While standard LLMs primarily employ causal attention, MLLMs often incorporate full attention for specific tokens. We introduce a mask efficiency factor $\eta_k$ to represent the additional computational overhead introduced by the full attention. For standard causal attention, $\eta_k$ is set to $0$, whereas for full attention, $\eta_k$ reflects the additional computations required. Thus, the computation cost for group $C_p$ is represented as:
        \begin{equation}{
        \small
        \begin{aligned}\mathcal{T}_{cp}(C_p) = \sum_{k} A_{k,p} (\alpha_1 (1+\eta_k) |s_k|^2+\alpha_2 |s_k|) + \beta_1,\label{eq:comp_mllm}\end{aligned}
        }\end{equation}
        where $\alpha_1, \alpha_2, \beta_1$ are profiled coefficients, and $\eta_k$ can be determined by the shape of attention mask.

        (2) \textbf{Communication Cost.} Ring Attention~\cite{brandon2023stripedattentionfasterring} utilizes point-to-point communication in a circular topology to exchange Key-Value (KV) blocks. The total communication cost is proportional to the sequence length $s_k$. The total communication cost is estimated as:
        \begin{equation}
        {
        \begin{aligned}
        \mathcal{T}_{cm}(C_p) = \frac{1}{ v_p} \sum_{k} A_{k,p} \alpha_3 |s_k| + \beta_2,
            \label{eq:comm}
        \end{aligned}
        }
        \end{equation}
        where $v_p$ is the point-to-point bandwidth within the CP group. 
        
        (3) \textbf{Total Time Estimation.} As the ring attention allows the communication cost and computation cost of attention to overlap, the communication time of attention needs to be excluded here. For simplicity, we do not detail the formula of the attention computation time $\mathcal{T}_{cpa}$ and communication time $\mathcal{T}_{cma}$, which has the similar form as Eq. \eqref{eq:comp_mllm} and \eqref{eq:comm}. 
        Then, the overall execution time is:
        \begin{equation}
        {
        \begin{aligned}
        \mathcal{T}(C_p) = \mathcal{T}_{cp}+\mathcal{T}_{cm}- \min (\mathcal{T}_{cpa}, \mathcal{T}_{cma}).
            \label{eq:time_cost}
        \end{aligned}
        }
        \end{equation}

        \subsection{Polynomial-Time Problem Solving}

        To efficiently solve the NP-hard problem defined in Eq.~\eqref{eq:minimize_time}, we propose a two-stage approximation algorithm: Memory-aware Sequence Packing followed by Dynamic Programming Resource Allocation. This significantly reduces the computational complexity and is able to find an approximate optimal solution while ensuring hardware constraints are strictly met.
        
        \textbf{Stage 1: Atomic Sequence Grouping via Best-Fit Decreasing (BFD). }
        To mitigate memory fragmentation and reduce the number of decision variables, we first group heterogeneous sequences into \textit{atomic sequence groups}. We sort all sequences in $\mathcal{B}$ in descending order of their memory requirements. For each long sequence, we determine its minimum required CP degree $d_{\min, k'} = \lceil \mathcal{M}(s_k) / E \rceil$, effectively initializing a ``bin'' with capacity $d_{\min, k'} \cdot E$. We then employ a BFD strategy to greedily pack shorter sequences into the remaining memory headroom of these established bins. This transforms the original set of $K$ sequences into $K'$ atomic groups $\{\mathcal{G}_1, \dots, \mathcal{G}_{K'}\}$, where $K' \leq K$. Each group $\mathcal{G}_k$ is treated as a single scheduling unit that requires at least $d_{\min, k'}$ ranks to satisfy memory constraint. Note that this successfully avoids communication redundancy caused by packing massive short sequences.
        
        \paragraph{Stage 2: Optimal Resource Assignment via 2D-Dynamic Programming.} 
        Given the atomic groups, we propose a 2D dynamic programming approach to determine the optimal CP degree $d_p$ and assignment $A_{k,p}$ for each group. Let $DP[i][j]$ denote the minimum achievable \textit{makespan} (i.e., the maximum execution time among all groups) for the first $i$ atomic groups utilizing a total of $j$ ranks. The state transition is defined as:
        \begin{equation}
        \small
        DP[i][j] = \min_{d \in [d_{\min, i}, j - d']}  \max ( DP[i-1][j-d], \\
        \mathcal{T}(\mathcal{G}_i, d) )
        \label{eq:dp_transition}
        \end{equation}
        where $d' = \sum_{m=1}^{i-1} d_{\min, m}$ ensures that sufficient ranks are reserved for the preceding $i-1$ groups. The function $\mathcal{T}(\mathcal{G}_i, d)$ represents the estimated execution time of group $\mathcal{G}_i$ under CP degree $d$. The inner $\max(\cdot)$ term identifies the execution bottleneck between the current group $\mathcal{G}_i$ and the previously allocated $i-1$ groups. The outer $\min(\cdot)$ then explores all valid CP degrees $d$ for the current group to find the smallest overall makespan.
        
        After populating the DP table, we \textbf{backtrack} from $DP[K'][N]$ to retrieve the optimal assignment matrix $\boldsymbol{A}$ and the corresponding CP degrees $\{d_p\}_{p=1}^P$. The total time complexity of this stage is $O(K'N^2)$ with only millisecond-level overhead, which can be overlapped with any sequence computation time through CPU next batch computation. The detailed algorithm is in Appendix~\ref{appendix:algorithm}.
        
        \section{Overall Workflow and Implementation}
        \setlength{\columnsep}{8pt}
        \begin{wrapfigure}{r}{0.55\linewidth}
        \begin{center}
        \vspace{-0.2cm}
        \includegraphics[width=0.55\textwidth]{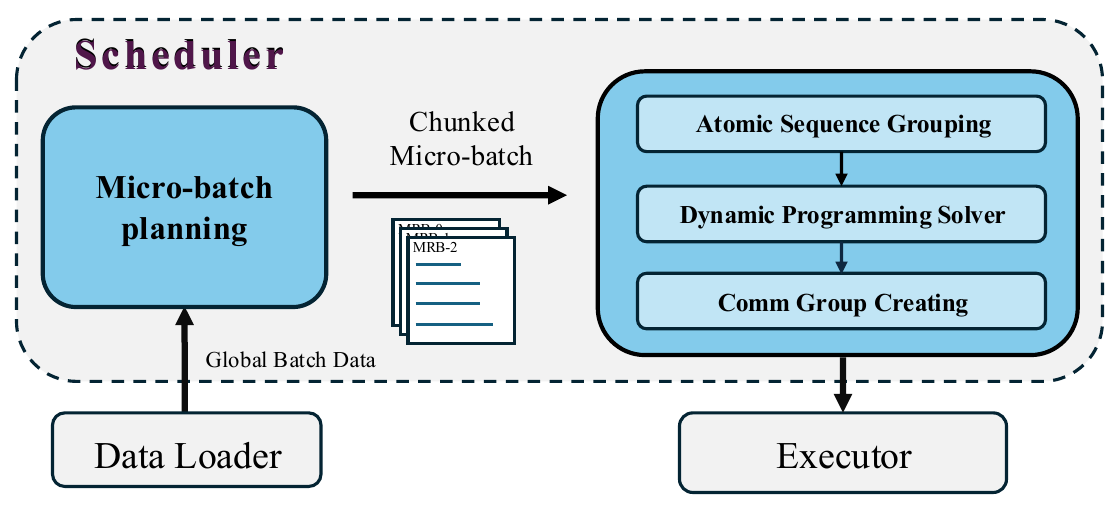}
        \end{center}
        \vspace{-0.25cm}
        \caption{Overall workflow of FCP. }
        \vspace{-0.2cm}
        \label{fig:wf}
        \end{wrapfigure}
        \textbf{Workflow.} The overall workflow of the FCP is illustrated in Figure~\ref{fig:wf}. (1) Given a training Global Batch, we first use \textbf{micro-batch planner} to chunk it into multiple micro-batches. (2) The scheduler first performs \textbf{memory-aware sequence packing} on micro-batch $\mathcal{B}$ to transform $K$ individual sequences into $K'$ atomic groups $\{\mathcal{G}_i\}$. (3) Subsequently, the scheduler invokes the \textbf{2D dynamic programming allocator} to determine the optimal assignment matrix $\boldsymbol{A}$ and CP degrees $\{d_p\}$ for each group. (4) After the DP solver is complete, the \textbf{executor} reallocates the hybrid parallel groups and dispatches data to each group.
        
        \textbf{Implementation Details.} 
        Our implementation is based on the Megatron-LM framework. For seamless integration with Megatron, all FCP components are implemented as add-on modules without modifying the Megatron core codebase. Our system primarily consists of two new core classes: Scheduler and Profiler, which together achieve robust dynamic parallelism. Below, we will detail several key aspects of the FCP implementation:

        1. \textbf{Decoupling Scheduling and Training.} In FCP, the workflow of each training batch consists of two phases: (1) the scheduling phase, which determines the optimal parallel strategy; and (2) the execution phase, which uses the allocated strategy to perform the real computation. To hide scheduling latency, while the device is computing the current batch, the CPU simultaneously solves the optimal plan for the next batch and prepares the necessary communication groups.

        2. \textbf{Profiler Integration and Cost Modeling.} Before the entire training process begins, the Profiler collects execution times for different sequence lengths and CP degrees, then establishes a functional relationship between runtime and factors (sequence length and CP degree), storing the corresponding parameters in the Profiler class. During the training, Profiler predicts the cost based on the stored parameters, providing the solver with a fast response.
        
        3. \textbf{Integration with Megatron Parallel Utilities.} FCP leverages Megatron's Parallel State Manager (MPU). To execute the latest dynamic parallel plan, The Scheduler class dynamically updates the CP parallel communication group within the MPU during each batch execution. Only the CP communication group is modified; all other forms of parallelism continue to run according to the original grid. This allows FCP to integrate seamlessly with existing Megatron frameworks.
        
        \section{Experiments}
        \subsection{Experimental Setup}
        
        \textbf{Hardware Environment.} All experiments are conducted on a cluster of 4 nodes. Each node is equipped with 16 Ascend 910C NPUs (64GB memory each), which are interconnected via HCCS within the node.
        
        \textbf{Baselines.} Given that all other dynamic parallelism baselines are closed-source, we follow prior work on distributed LLM training systems and compare FCP against two open-source state-of-the-art frameworks: DeepSpeed~\cite{Deepspeed} and Megatron-LM~\cite{narayanan2021efficient}. Megatron-LM supports 4D parallelism, integrating TP, PP, DP, and CP. DeepSpeed supports Ulysses-style sequence parallelism. 
        
        \textbf{Datasets.} We evaluated various baselines under a workload with a maximum sequence length of 128k. For LLMs, we use Qwen3 and Llama3/Llama3.2 models on three text datasets with different sequence-length distributions: GitHub~\cite{kocetkov2022stack3tbpermissively}, CommonCrawl~\cite{raffel2023exploringlimitstransferlearning}, and Wikipedia~\cite{wikidump}. For MLLMs, we use InternVL3 and Qwen3VL models on three video-language datasets: MSRVTT~\cite{xu2016msrvtt}, InternVid~\cite{wang2024internvidlargescalevideotextdataset}, and OpenVid~\cite{nan2025openvidm}. Further details are provided in Appendix~\ref{appendix:dtails}.
        
        \textbf{Evaluation Protocol.} For each baseline method, we use standard packing and tune the parallelism hyperparameters and select the best-performing configuration.  Since FCP adaptively schedules parallelism parameters, no manual per-run CP specification is required. To ensure a fair comparison, we fix the global batch size to 512 in all experiments. The evaluation metrics are the end-to-end training iteration time and the token throughput per device. During the performance test, we first warm up for 5 training steps, then record the average of the subsequent 10 steps for both metrics.
        
        \subsection{Evaluation}
        \begin{figure*}[h!]
          \centering
          \includegraphics[width=1.0\textwidth]{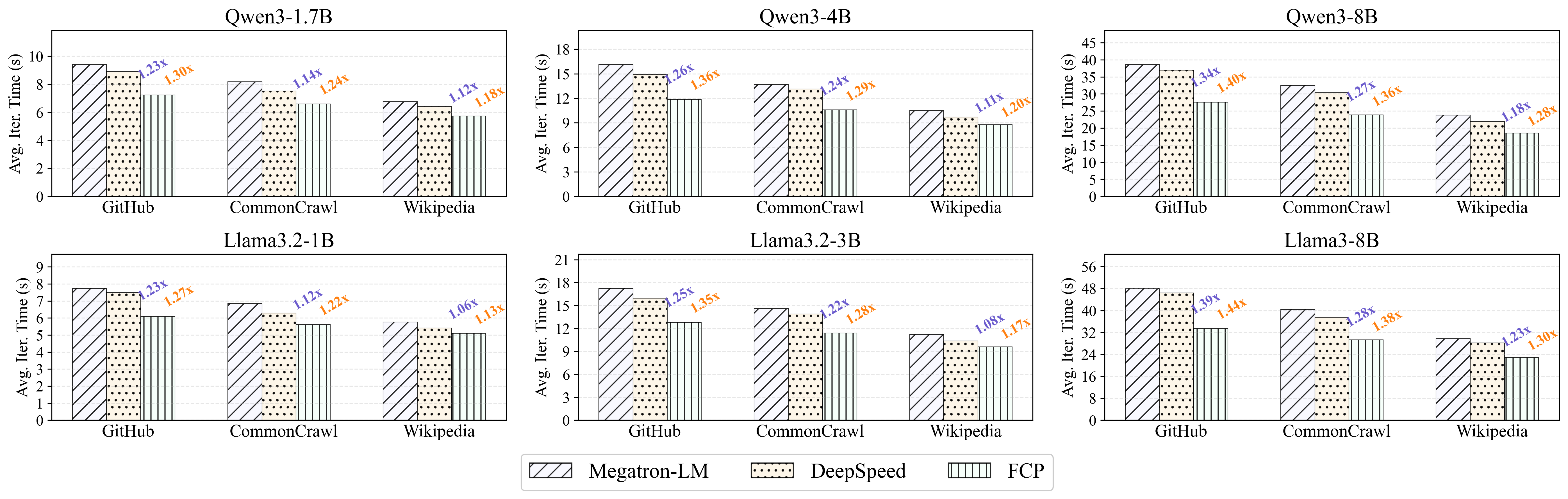}
          \caption{Average iteration time (in seconds) comparison on text datasets for Qwen3-1.7B/4B/8B, Llama3.2-1B/3B, and Llama3-8B. We compare Megatron-LM (diagonal stripes), DeepSpeed (dots), and FCP (vertical stripes) on GitHub, CommonCrawl, and Wikipedia.}
          \label{fig:text-model-comparison}
        \end{figure*}
        \textbf{LLM Training Performance.} To compare the end-to-end performance of FCP against the baselines on text-only LLM workloads, we conduct training on three text datasets (GitHub, CommonCrawl, and Wikipedia) using Qwen3 and Llama-family models of varying sizes (1B, 1.7B, 3B, 4B, and 8B). As shown in Figure~\ref{fig:text-model-comparison}, FCP consistently outperforms the best baseline (DeepSpeed or Megatron-LM) across all configurations, achieving speedups ranging from 1.05$\times$ (Llama3.2-1B on Wikipedia) to 1.37$\times$ (Qwen3-8B on GitHub). The improvement is particularly pronounced on long-tail text corpora, where FCP attains 1.37$\times$ and 1.36$\times$ speedups on GitHub with Qwen3-8B and Llama3-8B, respectively, and maintains robust gains on CommonCrawl (e.g., 1.30$\times$ for Qwen3-8B and 1.29$\times$ for Llama3-8B). Notably, FCP delivers speedups exceeding 1.2$\times$ for 8 out of 18 configurations, with the highest accelerations observed on larger models and datasets with heavier long-tail sequence distributions. These results demonstrate that FCP effectively improves text-only LLM training while scaling to larger model architectures.
        
        \begin{figure*}[h!]
          \centering
          \includegraphics[width=1.0\textwidth]{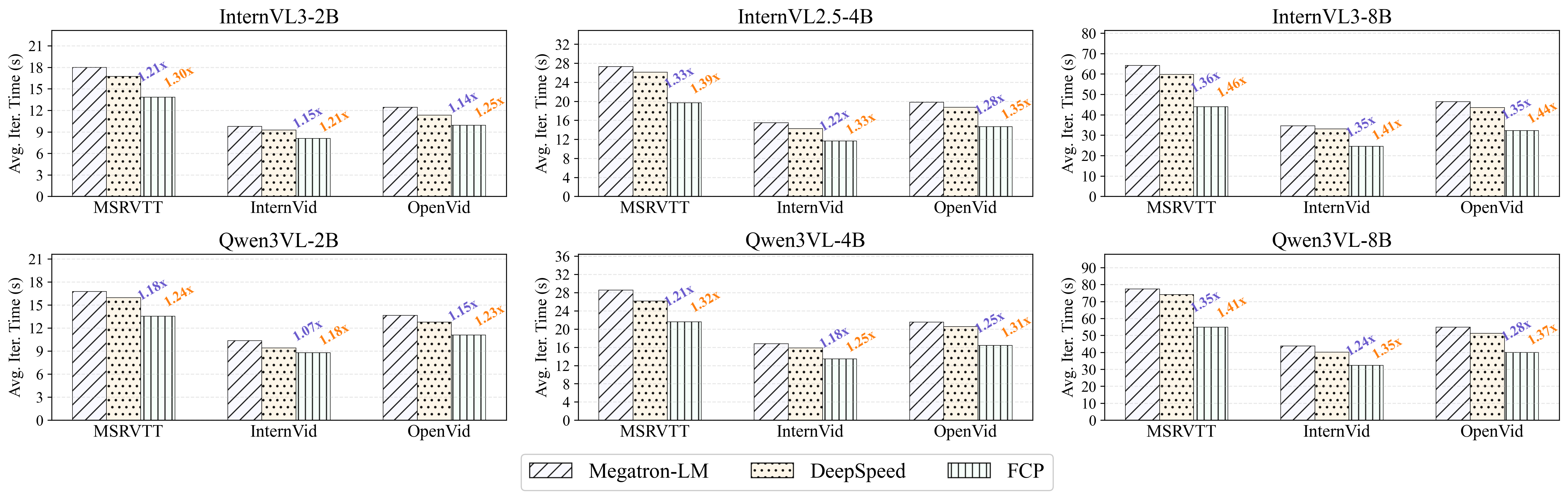}
          \caption{Average iteration time (in seconds) comparison for unfrozen MLLM training across MSRVTT, InternVid, and OpenVid datasets. The evaluated models include InternVL3-2B/8B, InternVL2.5-4B and Qwen3VL-2B/4B/8B.}
          \label{fig:unfrozen-mllm-comparison}
        \end{figure*}
        \textbf{Unfrozen MLLM Training Performance.} We further compare the end-to-end performance of FCP against the baselines on unfrozen MLLM training, where the vision encoder, projector, and LLM are trained jointly. We conduct experiments on three multimodal datasets (MSRVTT, InternVid, and OpenVid) using InternVL3 and Qwen3VL models of varying sizes (2B, 4B, and 8B). As shown in Figure~\ref{fig:unfrozen-mllm-comparison}, FCP consistently outperforms the best baseline across all configurations, achieving speedups ranging from 1.14$\times$ (InternVL3-2B on InternVid) to 1.38$\times$ (InternVL3-8B on MSRVTT). The improvement is particularly pronounced on larger 8B models, where FCP reaches 1.38$\times$ and 1.35$\times$ speedups for InternVL3-8B on MSRVTT and OpenVid, respectively, and also maintains strong gains for Qwen3VL-8B (1.37$\times$ on MSRVTT and 1.34$\times$ on OpenVid). Notably, FCP delivers speedups exceeding 1.2$\times$ for 15 out of 18 configurations, showing stable benefits under the heterogeneous computation of unfrozen MLLM training. These results demonstrate FCP's effectiveness across diverse multimodal scenarios while scaling to larger model architectures.
        
        
        \begin{wrapfigure}{r}{0.55\linewidth}
        \begin{center}
        \includegraphics[width=0.55\textwidth]{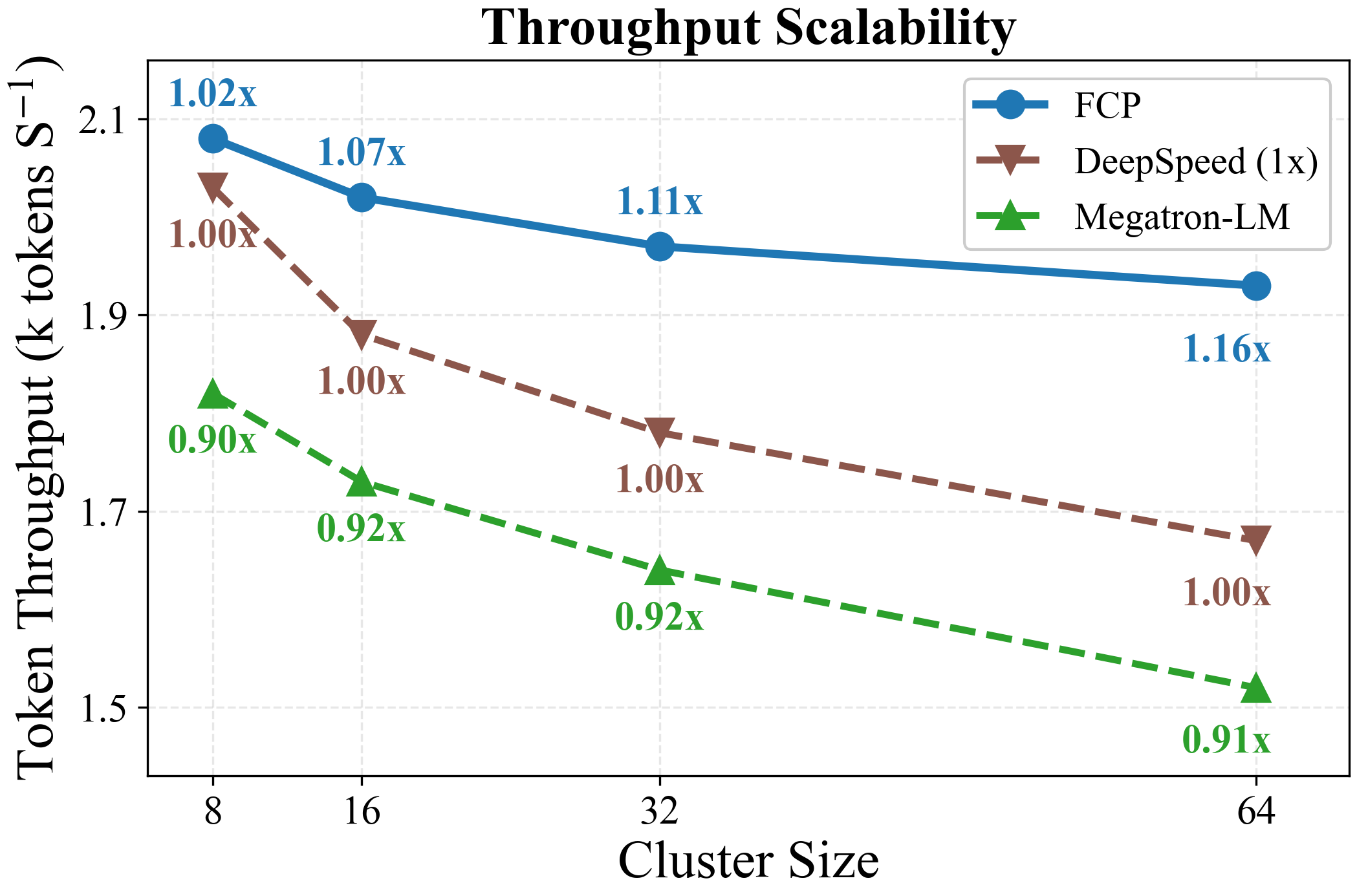}
        \end{center}
        \vspace{-0.2cm}
        \caption{ Token throughput (in k tokens/s) comparison across different cluster size (8, 16, 32, and 64) for FCP, DeepSpeed, and Megatron-LM.}
        \vspace{-0.25cm}
        \label{fig:scalability}
        \end{wrapfigure}
        \textbf{Scalability Analysis.} In this section, we analyze the performance of FCP on training clusters of different sizes. We run experiments on 1/2, 1, 2, and 4 nodes (corresponding to 8, 16, 32, and 64 devices respectively) and record the token throughput. As shown in Figure~\ref{fig:scalability}, The results show that
        as the cluster scale increases, while per-device throughput
        decreases due to growing communication overhead, FCP’s
        flexible and dynamic scheduling strategy mitigates this reduction effectively. Specifically, when scaling from 1/2 node
        (8 devices) to 4 nodes (64 devices), FCP maintains and even
        improves its relative throughput against DeepSpeed (1x)
        increasing from 1.02x to 1.16x. In contrast, DeepSpeed’s
        absolute throughput drops from approximately 2.0 to 1.7 k
        tokens/s, and Megatron-LM’s throughput declines from 1.8 to 1.52 k tokens/s. This demonstrates that FCP achieves superior scalability by minimizing the negative impact of communication overhead as the cluster size expands.
        
        \subsection{Time Analysis}
         \textbf{Scheduling Overhead.} Our proposed method uses an efficient solver with low time complexity. To validate this in real scenarios, we record empirical scheduling time, solver time, and the shortest GBS computation time. The tables below summarize the results under varying global batch sizes and cluster size when training Qwen3VL-2B.
        
        \begin{center}
        \small
        \begin{minipage}{0.48\linewidth}
        \centering
        \refstepcounter{table}\label{tab:time_consumption_gbs}
        \textbf{Table~\thetable.} Varying global batch size. \\
        [0.45em]
        \begin{tabular}{l ccc}
        \toprule
        & \multicolumn{3}{c}{GBS} \\
        \cmidrule(lr){2-4}
        & 128 & 256 & 512 \\
        \midrule
        Compute (s) & 6.34 & 7.58 & 11.65 \\
        Schedule (ms) & 448 & 696 & 905 \\
        Solver (ms) & 23 & 45 & 86 \\
        \bottomrule
        \end{tabular}
        \end{minipage}
        \hfill
        \begin{minipage}{0.48\linewidth}
        \centering
        \refstepcounter{table}\label{tab:time_consumption_npu}
        \textbf{Table~\thetable.} Varying Device count. \\
        [0.45em]
        \begin{tabular}{l ccc}
        \toprule
        & \multicolumn{3}{c}{Device Number} \\
        \cmidrule(lr){2-4}
        & 16 & 32 & 64 \\
        \midrule
        Compute (s) & 34.45 & 18.26 & 11.65 \\
        Schedule (ms) & 282 & 537 & 905 \\
        Solver (ms) & 20 & 44 & 86 \\
        \bottomrule
        \end{tabular}
        \end{minipage}
        \end{center}
        
        As shown in Tables~\ref{tab:time_consumption_gbs} and~\ref{tab:time_consumption_npu}, the solver time remains below 86 ms at the largest tested configuration. The overall scheduling time is also consistently shorter than a single GBS computation time when either increasing the GBS or scaling the cluster size. For example, with GBS=512 on 64 devices, scheduling takes 905 ms while computation takes 11.65 s. This small overhead allows scheduling to be overlapped with computation in practice.

        \textbf{Communication Group Management Overhead.}
        We also measure the overhead of creating and deleting CP communication groups, since FCP dynamically updates the CP group assignment across batches. The measured group-management overhead is around 1 ms in our implementation, which is negligible compared with the per-step computation time and therefore does not slow down the training process.
        
        \subsection{Profiler Error Analysis}
        
        To evaluate the accuracy of our profiler and cost estimator, we measure the prediction error across different modules and computational processes, then report the average error percentage.
        
        \begin{wraptable}{r}{0.49\linewidth}
        \centering
        \scriptsize
        \setlength{\tabcolsep}{3pt}
        \caption{Time cost estimation error (\%).}
        \label{tab:estimation_error}
        \vspace{-0.1cm}
        \begin{tabular}{l S[table-format=2.2] S[table-format=2.2] S[table-format=2.2]}
        \toprule
        {Model} & \multicolumn{3}{c}{Model Parameters} \\
        \cmidrule(lr){2-4}
                & {2B} & {4B} & {8B} \\
        \midrule
        Qwen3VL   & 7.93 & 6.71 & 4.27 \\
        InternVL2.5/3 & 7.48 & 6.54 & 4.12 \\
        \bottomrule
        \end{tabular}
        \vspace{-0.35cm}
        \end{wraptable}
        
        Table~\ref{tab:estimation_error} shows that the estimation error remains below 8\% across Qwen3VL and InternVL2.5/3 models. This level of accuracy is sufficient for scheduling as FCP mainly relies on the relative ordering of candidate CP strategies rather than exact absolute latency prediction.
        
        \subsubsection{Robustness to Profiler Noise}
        
        To further study whether profiler errors can change the final scheduling decision, we inject Gaussian noise into each coefficient of the cost model, with standard deviations ranging from 5\% to 50\% of the original coefficient. Then, we regenerate scheduling strategies with the perturbed models and evaluate them on real clusters.
        
        \begin{center}
        \small
        \begin{minipage}{0.48\linewidth}
        \centering
        \refstepcounter{table}\label{tab:robustness_32}
        \textbf{Table~\thetable.} Profiler noise on 32 Devs. \\[0.45em] 
        \begin{tabular}{lcc}
        \toprule
        Noise (\%) & Time/Step & CP Match (\%) \\
        \midrule
        0 (origin) & -- & 100.0 \\
        5  & +0.00\% & 100.0 \\
        10 & +0.00\% & 100.0 \\
        20 & +0.00\% & 100.0 \\
        30 & +3.27\% & 92.7 \\
        50 & +4.84\% & 91.3 \\
        \bottomrule
        \end{tabular}
        \end{minipage}
        \hfill
        \begin{minipage}{0.48\linewidth}
        \centering
        \refstepcounter{table}\label{tab:robustness_64}
        \textbf{Table~\thetable.} Profiler noise on 64 Devs. \\[0.45em]
        \begin{tabular}{lcc}
        \toprule
        Noise (\%) & Time/Step & CP Match (\%) \\
        \midrule
        0 (origin) & -- & 100.0 \\
        5  & +0.00\% & 100.0 \\
        10 & +0.00\% & 100.0 \\
        20 & +0.00\% & 100.0 \\
        30 & +5.73\% & 90.4 \\
        50 & +6.16\% & 88.8 \\
        \bottomrule
        \end{tabular}
        \end{minipage}
        \end{center}
        
        Tables~\ref{tab:robustness_32} and~\ref{tab:robustness_64} show that FCP is robust to realistic profiler noise: up to 20\% noise, the generated CP strategies are identical to the original strategy and the measured time per step does not change. Even under 50\% noise, the CP match rate remains around 90\%, and the overhead is only 4.84\% on 32 devices and 6.16\% on 64 devices.
        
        The robustness comes from the structure of the cost model. The quadratic term captures the attention cost, while linear terms capture linear-layer computation and communication. These components are positive, and as the sequence length increases the cost is increasingly dominated by the quadratic term. Therefore, moderate coefficient perturbations rarely change the relative ordering of candidate sequence partitions, which is the key information required by the dynamic-programming solver.
        
        \subsubsection{Case Study}\label{appendix:case_study}
        
        \begin{wraptable}{r}{0.48\linewidth}
        \vspace{-0.2cm}
        \centering
        \footnotesize
        \setlength{\tabcolsep}{4pt}
        \renewcommand{\arraystretch}{1.12}
        \caption{CP groups assigned by FCP.}
        \label{tab:case_study}
        \vspace{-0.05cm}
        \begin{tabular}{l|c|c}
        \hline
        Case Study & Case 1 & Case 2 \\
        \hline
        DeepSpeed & $\langle 8 \rangle \times 4$ & $\langle 4 \rangle \times 8$ \\
        \hline
        Megatron-LM & $\langle 8 \rangle \times 4$ & $\langle 4 \rangle \times 8$ \\
        \hline
        FCP &
        \begin{tabular}{@{}l@{}}
        $\langle 8 \rangle \times 1$ \\
        $\langle 6 \rangle \times 2$ \\
        $\langle 4 \rangle \times 1$ \\
        $\langle 2 \rangle \times 2$ \\
        $\langle 1 \rangle \times 4$
        \end{tabular} &
        \begin{tabular}{@{}l@{}}
        $\langle 4 \rangle \times 2$ \\
        $\langle 3 \rangle \times 4$ \\
        $\langle 2 \rangle \times 6$
        \end{tabular} \\
        \hline
        \end{tabular}
        \vspace{-0.2cm}
        \end{wraptable}
        
        To concretely illustrate FCP's dynamic partitioning of communication groups under varying data distributions, we present two actual parallel strategies employed within a single global batch. Table~\ref{tab:case_study} reports the heterogeneous CP groups used in the two cases, where each <$d$> $\times m$ indicates $m$ CP=$d$ groups. Case 1 is derived from OpenVid while Case 2 comes from MSRVTT. By mitigating communication redundancy and load imbalance introduced by static CP, FCP achieves speedups of $1.27\times$ in Case 1 and $1.19\times$ in Case 2.
        
        Furthermore, we conducted experiments under some rather extreme data distributions. When computing a batch with an extremely long (128k) sequence and many medium-to-short sequences (2-8k), our speedup even reached 2.24 $\times$.

        \newcommand{\mbname}{parallelism planner\xspace}
        \newcommand{\Mbname}{Parallelism Planner\xspace}
        \newcommand{\gbname}{sequence blaster\xspace}
        \newcommand{\Gbname}{Sequence Blaster\xspace}

        \section{Conclusion}
        In this paper, we presented Flexible
        Context Parallelism (FCP), an adaptive framework designed to optimize LLM training on heterogeneous datasets. By enabling non-power-of-two parallelism degrees and employing a two-stage approximation algorithm, FCP effectively eliminates load imbalance and redundant communication with millisecond
        level scheduling overhead. Our experimental results demonstrate that FCP achieves up to a 1.47× speedup in training
        throughput over Megatron-LM and DeepSpeed.  One limitation is that the performance gain will diminish with balanced data.

        \bibliography{reference}
        \bibliographystyle{plain}

        \newpage
        \appendix
        \onecolumn
        
        \section{Details of Experimental Setups}\label{appendix:dtails}
        
        \subsection{Dataset Details}\label{appendix:dataset}\label{appendix:text_setup}
        
        \textbf{Text datasets.} We use three text corpora with different sequence-length distributions to evaluate FCP under text-only long-context training.
        \begin{itemize}
            \item \textbf{GitHub} contains code-oriented documents with a pronounced long-tail distribution in sequence length. It stresses the ability of FCP to avoid over-provisioning CP groups for short samples while still supporting very long samples. A commonly used public source is The Stack on HuggingFace: \url{https://huggingface.co/datasets/bigcode/the-stack}.
            \item \textbf{CommonCrawl} contains large-scale web text and has a broad but less code-specific length distribution. We use it to represent general web-scale pre-training data. One public entry point is C4: \url{https://huggingface.co/datasets/allenai/c4}.
            \item \textbf{Wikipedia} contains encyclopedic articles with a shorter and denser length distribution than GitHub and CommonCrawl. We use it as a comparatively regular text corpus. A public version is available at \url{https://huggingface.co/datasets/wikimedia/wikipedia}.
        \end{itemize}
        
        \textbf{Multimodal datasets.} We use three video-language datasets with different duration and visual-token distributions to evaluate FCP under multimodal long-context training.
        \begin{itemize}
            \item \textbf{MSRVTT} consists of 10,000 video clips paired with 200,000 natural language descriptions from 20 diverse categories, measuring the video-to-text generation and multimodal understanding capabilities of machine learning models. The dataset is available at \url{https://huggingface.co/datasets/friedrichor/MSR-VTT}.
            \item \textbf{InternVid} comprises 10 million video clips paired with generated high-quality captions sourced from publicly available web videos, facilitating large-scale video-language pre-training and representation learning. The dataset is available at \url{https://huggingface.co/datasets/OpenGVLab/InternVid}.
            \item \textbf{OpenVid} comprises a curated collection of high-aesthetic video clips with a minimum resolution of $512 \times 512$, tailored for research institutions to enhance visual clarity, facilitating high-fidelity text-to-video training and quality-tuning as a complement to existing datasets. The dataset is available at \url{https://huggingface.co/datasets/nkp37/OpenVid-1M}.
        \end{itemize}
        
        \subsection{Model Details}\label{appendix:model}
        
        \textbf{LLM models.} For LLM experiments, we evaluate Qwen3-1.7B, Qwen3-4B, Qwen3-8B, Llama3.2-1B, Llama3.2-3B, and Llama3-8B. The selected models cover small and medium LLM settings and allow us to test whether FCP scales with increasing language-model computation. Table~\ref{tab:llm_models} summarizes the LLM configurations used in the experiments.
        
        \begin{table}[h]
        \centering
        \caption{\textbf{LLM models for evaluation.}}
        \begin{tabular}{|c|c|c|c|c|c|}
        \hline
        Model & \#Layers & \#Heads & \#Groups & Hidden Dim & FFN Hidden Dim\\ \hline
        Qwen3-1.7B & 28 & 16 & 8 & 2048 & 6144\\ \hline
        Qwen3-4B & 36 & 32 & 8 & 2560 & 9728\\ \hline
        Qwen3-8B & 36 & 32 & 8 & 4096 & 12288\\ \hline
        Llama3.2-1B & 16 & 32 & 8 & 2048 & 8192\\ \hline
        Llama3.2-3B & 28 & 24 & 8 & 3072 & 8192\\ \hline
        Llama3-8B & 32 & 32 & 8 & 4096 & 14336\\ \hline
        \end{tabular}
        \label{tab:llm_models}
        \end{table}
        
        \textbf{MLLM models.} For MLLM experiments, we evaluate InternVL3 and Qwen3VL models at different scales. Table~\ref{tab:models} summarizes the model configurations used in the experiments.
        
        \begin{table}[h]
        \centering
        \caption{\textbf{MLLM models for evaluation.}}
        \begin{tabular}{|c|c|c|c|c|c|}
        \hline
        Model & \#Layers & \#Heads & \#Groups & Hidden Dim & Vision Hidden Dim\\ \hline
        InternVL3-2B & 28 & 12 & 2 & 1536 & 1024\\ \hline
        InternVL2.5-4B & 36 & 16 & 2 & 2048 & 1024 \\ \hline
        InternVL3-8B & 28 & 28 & 4 & 3584 & 1024\\ \hline
        Qwen3-VL-2B & 28 & 16 & 8 & 2048 & 1024 \\ \hline
        Qwen3-VL-4B & 36 & 32 & 8 & 2560 & 1024\\ \hline
        Qwen3-VL-8B & 36 & 32 & 8 & 4096 & 1152 \\ \hline
        \end{tabular}
        \label{tab:models}
        \end{table}
        
        \subsection{Baseline Details}\label{appendix:baseline}
        
        We use the same two static parallel training baselines for both LLM and MLLM experiments. For each baseline, we tune the parallelism hyperparameters and report the best-performing configuration under the same global batch size.
        \begin{itemize}
            \item \textbf{Deepspeed-Ulysses}~\cite{DeepSpeed-Ulysses} addresses the memory constraints imposed by extreme sequence lengths in LLM training through a sequence parallelism mechanism involving hybrid partitioning. In contrast to Megatron-SP~\cite{korthikanti2022reducingactivationrecomputationlarge}, which partitions the sequence dimension only for LayerNorm and Dropout while relying on ring-based communication for Self-Attention, DeepSpeed Ulysses partitions the input data along the sequence dimension across all modules. To enable global attention without storing the full sequence on a single device, it employs an all-to-all collective communication primitive to transform the data layout.
            \item \textbf{Megatron-LM}~\cite{shoeybi2020megatronlmtrainingmultibillionparameter} composes tensor, pipeline, and data parallelism (PTD-P) to scale language model training to trillions of parameters. To address the idle time inherent in standard pipeline parallelism, it proposes an interleaved pipelining schedule that assigns virtual model chunks to each pipeline stage.
        \end{itemize}

        \section{FCP Optimization Algorithm}\label{appendix:algorithm}

        \begin{algorithm2e}[h]
        \caption{2D-Dynamic Programming}
        \label{alg:dp_allocation}
        \scalebox{0.95}{
        \begin{minipage}{\textwidth}
        
        \KwIn{Groups $\mathcal{G}$, Min-degrees $\{d_{\min, k'}\}$, Total ranks $N$}
        \KwOut{CP degrees $\{d_p\}_{p=1}^P$}
        \text{Initialization}\;
        $K' \leftarrow |\mathcal{G}|, DP[K'+1][N+1], Path[K'+1][N+1] $\;
        $DP[0][0] \leftarrow 0$\;
        
        \For{$k = 1$ \KwTo $K'$}{
            $R_{remain} \leftarrow \sum_{z=k+1}^{K'} d_{\min, z}$\;
            
            \For{$j = (\sum_{i=1}^{k} d_{\min, i})$ \KwTo $(N - R_{remain})$}{
                \For{$d = d_{\min, k}$ \KwTo $(j - \sum_{i=1}^{k-1} d_{\min, i})$}{
                    $cost \leftarrow \max(DP[k-1][j-d], \mathcal{T}(G_k, d))$\;
                    
                    \If{$cost < DP[k][j]$}{
                        $DP[k][j] \leftarrow cost$\; 
                        $Path[k][j] \leftarrow d$\;
                    }
                }
            }
        }
        $p\leftarrow K'+1$\;
        $q\leftarrow N+1$\;
        \While{$p,q > 0$}{
            $d_p \leftarrow Path[p][q]$\; 
            $p \leftarrow p - 1, \ q \leftarrow q - d_p$\;
        }
        \Return{$\{d_1, \dots, d_P\}$}\;
        \end{minipage}
        }
        \end{algorithm2e}
        
        \section{Frozen MLLM Results}\label{appendix:frozen_mllm}
        
        \begin{figure*}[h!]
          \centering
          \includegraphics[width=1.0\textwidth]{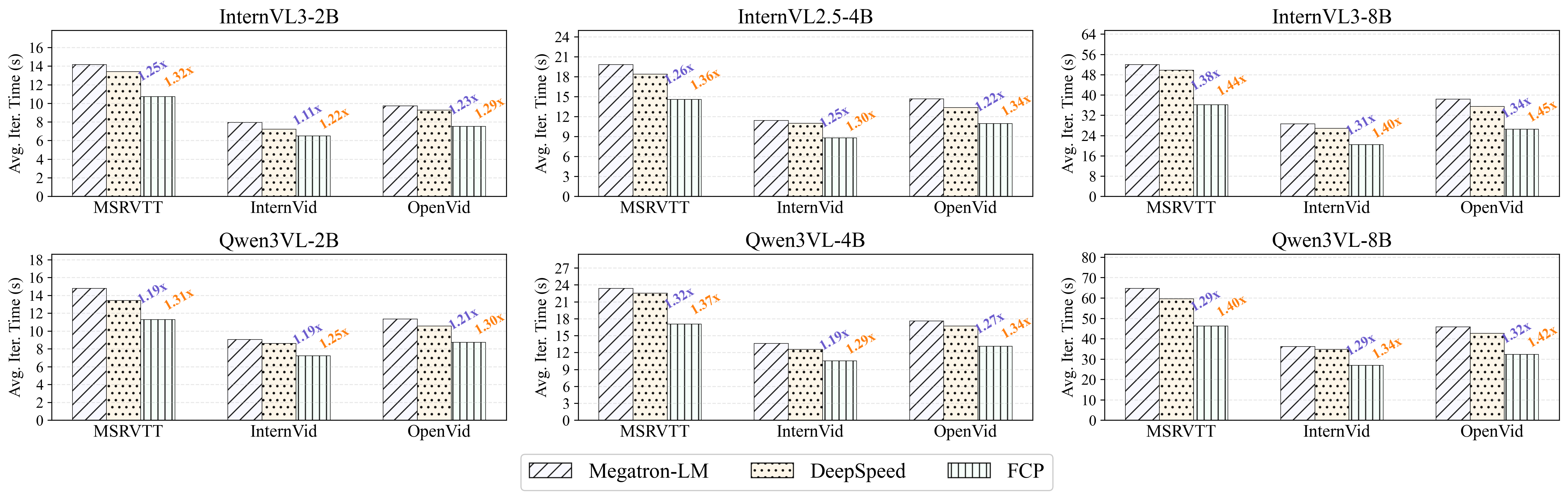}
          \caption{Average iteration time (in seconds) comparison for frozen-vision MLLM training across MSRVTT, InternVid, and OpenVid datasets. FCP keeps the same scheduling mechanism while the vision encoder is frozen, yielding shorter wall time than the unfrozen setting.}
          \label{fig:frozen-mllm-comparison}
        \end{figure*}
        
        \textbf{Frozen MLLM Training Performance.} We place the frozen-vision MLLM results in the appendix because the main paper focuses on text-only LLM training and the more challenging unfrozen MLLM setting. To supplement the main results, we evaluate frozen-vision MLLM training on the same three multimodal datasets (MSRVTT, InternVid, and OpenVid) using InternVL3, InternVL2.5, and Qwen3VL models of varying sizes. As shown in Figure~\ref{fig:frozen-mllm-comparison}, freezing the vision encoder reduces the overall wall time for all methods, while FCP still consistently outperforms the best static baseline across all configurations, achieving speedups ranging from 1.16$\times$ (InternVL3-2B on InternVid) to 1.40$\times$ (InternVL3-8B on MSRVTT). The improvement remains especially clear on larger 8B models, where FCP reaches 1.40$\times$ and 1.37$\times$ speedups for InternVL3-8B on MSRVTT and OpenVid, respectively, and also maintains strong gains for Qwen3VL-8B (1.39$\times$ on MSRVTT and 1.35$\times$ on OpenVid). Overall, FCP delivers speedups exceeding 1.2$\times$ for 15 out of 18 configurations, demonstrating that its dynamic CP scheduling remains effective even when the vision encoder is frozen and the training workload becomes lighter than the unfrozen setting.

        \end{document}